\documentclass{article}

\usepackage[english]{babel}
\usepackage{amsmath}
\usepackage{amsfonts}

\usepackage[letterpaper, a4paper, top=2cm,bottom=2cm,left=3cm,right=3cm,marginparwidth=1.8cm]{geometry}

\usepackage{amsmath}
\usepackage{graphicx}
\usepackage[colorlinks=true, allcolors=blue]{hyperref}
\usepackage{authblk}

\usepackage{tabu}                     
\usepackage{multirow}                
\usepackage{multicol}                 
\usepackage{multirow}                
\usepackage{float}                   
\usepackage{makecell}                 
\usepackage{booktabs}                
\providecommand{\keywords}[1]
{
  \small	
  \textbf{{Keywords:}} #1
}

\title{Towards Ground-truth-free Evaluation of Any Segmentation in Medical Images\thanks{Work in progress. A. Senbi and T. Huang contributed equally.}}
\author{Ahjol Senbi, Tianyu Huang, Fei Lyu, Qing Li, Yuhui Tao, Wei Shao, Qiang Chen, Chengyan Wang, Shuo Wang, Tao Zhou, Yizhe Zhang\thanks {A. Senbi, T. Huang, Q. Chen, T. Zhou and Y. Zhang are from the Nanjing University of Science and Technology, Nanjing, China. Q. Li, Y. Tao, C. Wang and S. Wang are from Fudan University, Shanghai, China. F. Lyu is from Hong Kong Baptist University, Hong Kong, SAR. W. Shao is from Nanjing University of Aeronautics and Astronautics, Nanjing, China.\\Email: yizhe.zhang.cs@gmail.com, shuowang@fudan.edu.cn}}
\date{}

\newtheorem{theorem}{Theorem}[section] 
\newtheorem{proposition}[theorem]{Proposition}

\newtheorem{definition}{Definition}[section] 
\begin{document}
\maketitle

\begin{abstract}
We explore the feasibility and potential of building a ground-truth-free evaluation model to assess the quality of segmentations generated by the Segment Anything Model (SAM) and its variants in medical imaging. This evaluation model estimates segmentation quality scores by analyzing the coherence and consistency between the input images and their corresponding segmentation predictions. Based on prior research, we frame the task of training this model as a regression problem within a supervised learning framework, using Dice scores (and optionally other metrics) along with mean squared error to compute the training loss. The model is trained utilizing a large collection of public datasets of medical images with segmentation predictions from SAM and its variants. We name this model \textbf{EvanySeg} (\textbf{Eva}luation of \textbf{Any} \textbf{Seg}mentation in Medical Images). Our exploration of convolution-based models (e.g., ResNet) and transformer-based models (e.g., ViT) suggested that ViT yields better performance for this task. EvanySeg can be employed for various tasks, including: (1) identifying poorly segmented samples by detecting low-percentile segmentation quality scores; (2) benchmarking segmentation models without ground truth by averaging quality scores across test samples; (3) alerting human experts to poor-quality segmentation predictions during human-AI collaboration by applying a threshold within the score space; and (4) selecting the best segmentation prediction for each test sample at test time when multiple segmentation models are available, by choosing the prediction with the highest quality score. Models and code will be made available at \url{https://github.com/ahjolsenbics/EvanySeg}.

\keywords{Ground-truth-free Segmentation Evaluation, Quality Assessment, Medical Image Segmentation, Foundation Model for Trustworthy Medical AI}


\end{abstract}

\section{Introduction}






Pioneering work by Robinson et al.~\cite{robinson2018real} demonstrated the feasibility of using segmentation maps alongside raw images to predict Dice coefficient scores for the segmentation quality of the left-ventricular cavity, left-ventricular myocardium, and right-ventricular cavity in cardiovascular magnetic resonance scans. Li et al.~\cite{li2022towards} then developed a reference-based method to evaluate both image-level and pixel-level segmentation quality. More recently, Wundram et al.~\cite{wundram2024conformal} introduced the use of conformal range prediction to estimate the upper and lower bounds of the Dice score in ground-truth-free segmentation evaluation. Chen et al.~\cite{chen2024quality} advanced the field further by employing vision-language models for segmentation quality assessment, incorporating text to specify the segmentation class under investigation (e.g., ``Spleen"). These developments underscore that segmentation quality assessment is increasingly recognized as a crucial step toward the reliable deployment of medical image segmentation models and the facilitation of effective human-AI collaboration in medical image analysis.

In parallel, the Segment Anything Model (SAM)~\cite{kirillov2023segment}, a class-agnostic segmentation model that utilizes points and/or bounding boxes as prompts for segmentation, has gained significant attention since its introduction. Many efforts have been made to adapt, fine-tune, and apply SAM in medical imaging. MedSAM~\cite{ma2024segment} and SAM-Med2D~\cite{cheng2023sam} both adopt the bounding box and/or point prompt scheme for generating segmentations in medical imaging contexts. Owing to their flexibility and utility, SAM and SAM variants have been widely employed in the field of medical image analysis. Given this context, it is both desirable and essential to have an external system capable of automatically and reliably assessing the quality of the segmentations generated by SAM models. Motivated by this, we propose the development of a segmentation quality evaluator, termed {$\textrm{EvanySeg}$}, designed to work alongside SAM models in medical image segmentation. Specifically, when SAM (e.g., the original SAM and/or SAM variants) generates a segmentation prediction, {$\textrm{EvanySeg}$} processes this output along with the input image and produces a score that accurately reflects the true segmentation quality. 
The highlights of this work can be summarized in the following three points.

\begin{itemize} 
\item We explore a new approach to enhance the effectiveness and reliability of segment-anything models (SAM) in medical imaging by developing a companion model, named $\textrm{EvanySeg}$. This model is specifically designed to evaluate the quality of segmentations produced by SAM and its variants, aiming to promote more accurate and dependable segmentation outcomes.

\item Our theoretical analysis suggests that developing a segmentation evaluation model is indeed feasible. In many instances, this task is as manageable as, or even simpler than, creating the segmentation model itself. We deliberately designed the model and training procedure to be simple and straightforward, striving for ease of implementation and use in practice.

\item We trained the segmentation evaluation model using a large collection of public available medical imaging data and tested its performance across various practical applications utilizing both public and private datasets. The results demonstrate the model's effectiveness and utility in estimating segmentation quality for medical images without the need for ground truth masks. We will make the code and trained models publicly available to foster and advance research for reliable medical segmentation AI system. 

\end{itemize}


\section{Related Work}

\subsection{Segmentation Quality Assessment}

Huang et al.~\cite{huang2016qualitynet} introduced a learning-based approach for estimating segmentation quality using deep learning (DL) networks. They developed three network configurations, focusing primarily on how segmentation masks are integrated with features or images during the estimation of the segmentation quality (SQ) score. Zhou et al.~\cite{zhou2020robust} proposed a sequential network architecture for segmentation quality assessment (SQA). Devries et al.~\cite{devries2018leveraging} employed uncertainty maps to assist in the estimation of quality scores by combining raw images, segmentation maps, and uncertainty maps, which were then input into a DL-based network for SQA. Rottmann et al.~\cite{rottmann2020prediction} suggested using the dispersion of softmax probabilities to predict the true segmentation intersection over union (IoU). Rahman et al.~\cite{rahman2022fsnet} introduced an encoder-decoder architecture designed to detect segmentation failures at the pixel level. They extracted multi-scale features from the segmentation model and fed them into the decoder to generate a map that highlights misclassified or mis-segmented pixels. Valindria et al.~\cite{valindria2017reverse} proposed Reverse Classification Accuracy (RCA) for SQA, which requires a reference database containing image and ground-truth map pairs. Finally, Robinson et al.~\cite{robinson2018real} developed convolutional neural networks (CNNs) to directly predict the Dice score for an image and segmentation map pair, utilizing a large dataset of images with ground-truth labels to train the CNN for this prediction task. Wang et al. ~\cite{wang2020deep} examined the latent distribution of image-segmentation pairs and proposed a generative method to identify surrogate ground truths. More Recently, Wundram et al.~\cite{wundram2024conformal} proposed to utilize conformal range prediction to predict the upper and lower bounds of the Dice score in ground-truth free segmentation evaluation. Chen et al.~\cite{chen2024quality} developed a visual-language model for estimating segmentation and label quality in medical segmentation datasets. This model was trained on over 4 million image-label pairs and can predict segmentation quality for 142 body structures in CT (Computed Tomography) images, showing good correlation with ground truth quality measures. 



\subsection{Segment Anything Models for Medical Images}
Ma et al.~\cite{ma2024segment} compiled 84 publicly available datasets comprising over one million images from 10 imaging modalities to train a Segment Anything Model specifically for medical images. Similarly, Cheng et al.~\cite{cheng2023sam} developed two versions of their medical SAMs, with one model tailored for 2D images and another for 3D images. More recently, Zhao et al.~\cite{zhao2023one} introduced a Segment Anything Model using text prompts (SAT model). This work involved the curation of an extensive dataset that includes over 11,000 3D medical image scans sourced from 31 segmentation datasets. Rigorous standardization was applied to both the visual scans and the label space during dataset construction. The SAT model leverages text queries in conjunction with input images to generate segmentation masks corresponding to the content specified in the text. Despite these advancements, there remain considerable performance gaps between the current medical SAMs and the level of segmentation accuracy required for clinical applications. Preliminary studies also suggest that the features learned by medical foundation models do not necessarily translate into significant performance gains when compared to much smaller, well-established models~\cite{alfasly2023foundation}.


\section{Materials and Methods}
\subsection{Preliminary}\label{sec:overview}
\begin{figure}[t] 
    \centering
    \includegraphics[width=1\linewidth]{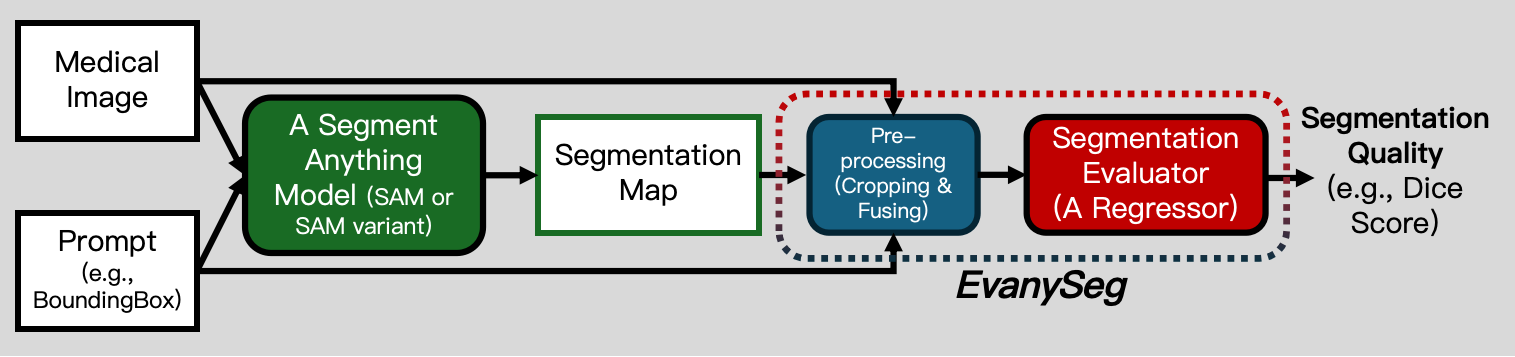}
    \caption{ {$\textrm{EvanySeg}$} is a companion model to SAM and its variants, designed to enhance reliability and trustworthiness in the deployment of SAM (and its variants) on medical images.}
    \label{fig:workflow}
\end{figure}

Given an image sample \( x \in \mathbb{R}^{w \times h \times C} \) and a prompt (e.g., a bounding box), a Segment Anything Model (SAM) generates a segmentation prediction map \( \hat{y} \in \mathbb\{0,1\}^{w \times h} \), where \( w \) represents the width and \( h \) the height of the image, and $C$ represents the number of channels of the image. To evaluate the segmentation \( \hat{y} \), the conventional approach is to obtain ground truth annotations \( y \) from human experts and compare \( \hat{y} \) with \( y \) using a predefined evaluation metric, such as the Dice score. Let \( \pi \) denote the evaluation metric; then, the sample-wise quality evaluation result is given by \( q = \pi(\hat{y}, y) \). In practice, it is highly desirable to predict segmentation quality without relying on ground truth annotations. Following the approach of~\cite{robinson2018real}, we aim to build a regression model, denoted as \( \tau \), that directly predicts \( q \) based on the segmentation prediction \( \hat{y} \) from SAM and the raw input image \( x \). The workflow of \(\textrm{EvanySeg}\), a companion model designed to work alongside SAM, is illustrated in Figure~\ref{fig:workflow}.

\subsection{Theoretical Analysis}   

\begin{quote} 
\centering 
``It may not know what a 100\% accurate segmentation is, but it can still tell, to some extent, whether a segmentation is good or bad.''
\end{quote}

\begin{theorem}
Given an image, estimating the segmentation quality (e.g., Dice score) of a segmentation map (Problem A) for this image is no harder than generating a perfectly accurate segmentation map (Problem B) for this image.
\end{theorem}

\noindent\underline{Reducing Problem A to Problem B:} To solve Problem A (i.e., predicting the Dice score for a given segmentation map $\hat{y}$), one can first solve Problem B by generating the true segmentation map $y$. Once $y$ is available, calculating the Dice score between $\hat{y}$ and $y$ becomes a straightforward task. This approach involves solving only one instance of Problem B, followed by a simple Dice score computation.

\noindent\underline{Reducing Problem B to Problem A:} To solve Problem B (i.e., generating a perfectly accurate segmentation map), one could theoretically generate all possible segmentation maps for a given image $x$. Since the number of potential segmentation maps grows exponentially, there are a vast number of possibilities. For each candidate segmentation map, one could then solve Problem A (i.e., predicting the Dice score) to assess how close the generated map $\hat{y}$ is to the true segmentation map $y$. The optimal segmentation map $\hat{y}$ would be the one that maximizes the predicted Dice score. Thus, solving Problem B in this manner would require solving an exponential number of instances of Problem A. See Theorem~\ref{theorem2} for a more efficient way to reduce problem B to problem A, resulting in significantly fewer—though still many—instances of problem A that need to be solved, rather than an exponentially large number.

\noindent\underline{In summary:} Since solving Problem A requires only one instance of Problem B followed by minimal computational effort, Problem B is at least as hard as Problem A (Problem A is no harder than Problem B). Furthermore, because solving Problem B can require solving many instances of Problem A, Problem B can be significantly harder in general. As a result, predicting the Dice score (Problem A) is \textbf{generally} easier than creating an accurate segmentation map (Problem B). Though, in the most challenging cases of Problem A, solving those difficult instances can be equivalent to solving Problem B.

\begin{theorem} \label{theorem2}
There exists a set of instances of Problem A, referred to as the core set, for which the complexity of solving this set is equivalent to that of solving Problem B. The size of the set is linearly proportional to the number of pixels in the image. 
\end{theorem}

We construct this set as follows. For simplicity of this analysis, we assume that the segmentation task is binary, with the ground truth \(y\) represented as a binary map, where \(1\) indicates the object of interest and \(0\) indicates the background or other objects. For each pixel in the segmentation map, we can create a pair of segmentation maps that differ only at that pixel. We then invoke the solver for Problem A on these two segmentation maps and compare the output values to determine whether the corresponding pixel should be \(1\) or \(0\) in the correct segmentation (the ground truth). By repeating this process for every pixel in the segmentation map, we can generate the ground truth segmentation (solving Problem B). The total number of instances of Problem A being solved in this process is precisely twice the total number of pixels in the image. This set of instances constitutes the core set of Problem A, where solving it leads to the resolution of Problem B and subsequently enables the resolution of the other instances of Problem A outside the core set (exponentially many).


\begin{definition}
Given an image \( x \), and any segmentation map \( \hat{y} \) of the image, a segmentation quality evaluation model \( \tau \) is considered ``absolutely accurate" if the following condition holds:
\[
\tau(\hat{y}, x) - \pi(\hat{y}, y) = 0
\],
where \( \pi(\hat{y}, y) \) is computed by a pre-defined metric $\pi$ that measures the quality of the segmentation map \( \hat{y} \) with respect to the ground truth \( y \).
\label{d1}
\end{definition}

\begin{definition}
Given any two images \( x_1 \) and \( x_2 \) (where \( x_1 \) can be the same as \( x_2 \)), and any two segmentation maps \( \hat{y}_1 \) for \( x_1 \) and \( \hat{y}_2 \) for \( x_2 \), a segmentation quality evaluation model \( \tau \) is considered ``relative accurate" if the following condition holds (assuming $\pi(\hat{y}_1, y_1) \neq \pi(\hat{y}_2, y_2)$):
\[
(\tau(\hat{y}_1, x_1) - \tau(\hat{y}_2, x_2)) \times (\pi(\hat{y}_1, y_1) - \pi(\hat{y}_2, y_2)) > 0
\],
where \( \pi(\hat{y}_1, y_1) \) and \( \pi(\hat{y}_2, y_2) \) are computed by a pre-defined metric $\pi$ that measures the quality of the segmentation maps \( \hat{y}_1 \) and \( \hat{y}_2 \) with respect to their ground truths \( y_1 \) and \( y_2 \).
\label{d2}
\end{definition}

\begin{proposition}
Achieving a segmentation quality evaluation model \( \tau \) that is relative accurate (Definition~\ref{d2}) is generally easier than achieving a model that is absolutely accurate (Definition~\ref{d1}).
\end{proposition}

The above is straightforward to show, since a model that is absolutely accurate is also relatively accurate, but not vice versa.

\begin{definition}
Given any two images \( x_1 \) and \( x_2 \) (where \( x_1 \) can be the same as \( x_2 \)), and any two segmentation maps \( \hat{y}_1 \) for \( x_1 \) and \( \hat{y}_2 \) for \( x_2 \), a segmentation quality evaluation model \( \tau \) is considered \(\beta\)-relative accurate if, whenever \( |\pi(\hat{y}_1, y_1) - \pi(\hat{y}_2, y_2)| \geq \beta \), the following condition holds (assuming $\pi(\hat{y}_1, y_1) \neq \pi(\hat{y}_2, y_2)$): \[ (\tau(\hat{y}_1, x_1) - \tau(\hat{y}_2, x_2)) \times (\pi(\hat{y}_1, y_1) - \pi(\hat{y}_2, y_2)) > 0 \], where \( \pi(\hat{y}_1, y_1) \) and \( \pi(\hat{y}_2, y_2) \) are metrics that measure the quality of the segmentation maps \( \hat{y}_1 \) and \( \hat{y}_2 \) with respect to their ground truths \( y_1 \) and \( y_2 \).
\label{d3}
\end{definition}

\begin{proposition}
Given \( \beta_1 < \beta_2 \), achieving a segmentation quality evaluation model \( \tau \) that is \(\beta_1\)-relative accurate (according to Definition~\ref{d3}) is more challenging than achieving a model that is \(\beta_2\)-relative accurate.\label{p3}
\end{proposition}

So far, we have established several key theoretical aspects in developing a segmentation quality evaluation model. Most notably, we have demonstrated that constructing such a model is not more difficult than creating the segmentation model itself, and in some cases, it may be easier. Furthermore, achieving a ``relative accurate" segmentation quality evaluation model is a more realistic and attainable goal compared to striving for an absolutely accurate evaluation model. Lastly, we introduced a parameter $\beta$ that further elucidates the complexity and difficulty involved in building the evaluation model, which will demonstrate its usefulness when analyzing empirical results in Section~\ref{exp_sec1}.

\subsection{Training Data}
Preparing the data for training {$\textrm{EvanySeg}$} is a critical step in this work. To build the evaluation model, which serves as a companion to the Segment Anything Model (SAM), we integrate SAM into the process of preparing the training data for {$\textrm{EvanySeg}$}. Several variants of SAM are utilized: the original release from Meta~\cite{kirillov2023segment}, the later-released MedSAM~\cite{ma2024segment}, and SAM-Med2D~\cite{cheng2023sam}. We employ these three SAM variants in the preparation of training data.

For each image $x_i$ where $i = 1, 2, \dots, n$, and for every prompt $p_{i,j}$, the ground truth mask is denoted as $y_{i,j} \in \{0,1\}^{w \times h}$, where $j = 1, 2, \dots, k_i$. Here, $w$ is the width and $h$ is the height of the image, and $k_i$ represents the number of objects in image $x_i$. Each object is specified with a dedicated prompt. Prompts can take the form of bounding boxes or center points. For an image $x_i$ with a prompt $p_{i,j}$, we apply a SAM variant to generate a segmentation output denoted as $\hat{y}_{i,j}$. We then compute a segmentation quality score $q_{i,j} = \pi (\hat{y}_{i,j}, y_{i,j})$, where $\pi$ may be as simple as the Dice Similarity Coefficient (DSC). This process is repeated for every image and every prompt. The training data for {$\textrm{EvanySeg}$} is thus prepared as $(x_i, p_{i,j}, y_{i,j}, \hat{y}_{i,j}, q_{i,j})$, where $i = 1, 2, \dots, n$, and $j = 1, 2, \dots, k_i$. This procedure is repeated for each SAM variant. We utilize a subset of the data curated in SA-Med2D-20M~\cite{ye2023sa} for constructing the training data for {$\textrm{EvanySeg}$}. More details on data preparation can be found in our code release.


\begin{figure}[t] 
    \centering
    \includegraphics[width=1\linewidth]{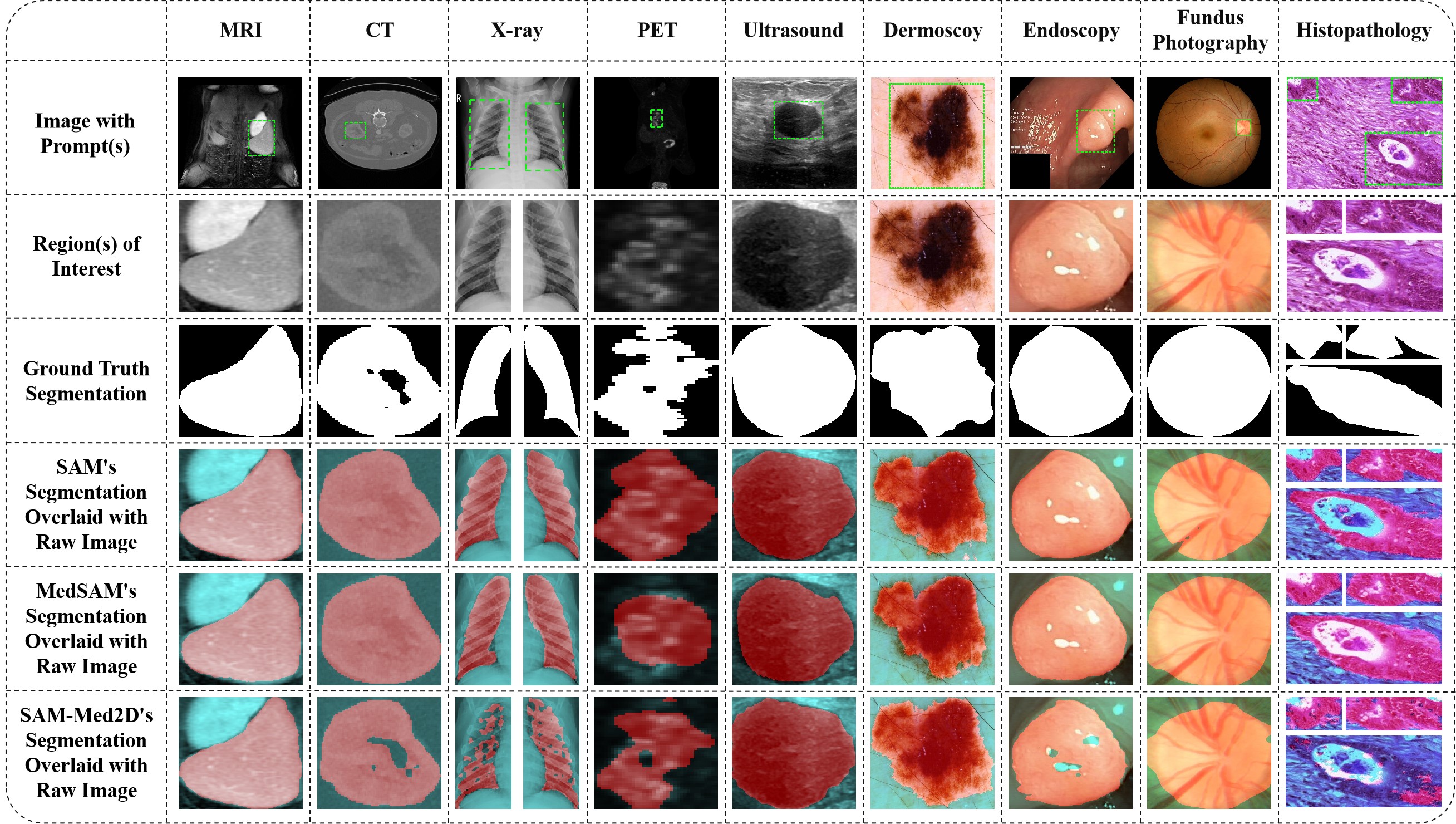}
    \caption{An illustration of the training data in a scenario where bounding box was used as prompt. The regions of interest are defined by the bounding box prompts.}
    \label{fig:data_demo}
\end{figure}

\subsection{Model Architecture}
{$\textrm{EvanySeg}$} consists of two components: the pre-processing model and the regression model. The pre-processing model aims to combine the segmentation map with the input image, and it uses the prompt to crop the regions of interest, constructing the inputs for the regression model. The regression model then takes the output from the pre-processing model and generates a score that reflects the segmentation quality. Detailed designs for the two models are provided below.

\underline{Pre-processing module:} Given an image \( x_i \) with a segmentation generated by SAM, denoted as \( \hat{y}_{i,j} \), our objective is to combine these two elements into a single compact map. If \( x_i \) is a gray-scale image, we expand its single channel to three channels; if \( x_i \) is a color image (with three channels), it remains unchanged. To merge the image and the segmentation, we add \( \hat{y}_{i,j} \) to the first channel (red channel) of \( x_i \) using the following formula: \( x_i [:,:,0] = 0.5 \times x_i [:,:,0] + 0.5 \times \hat{y}_{i,j} \). Although there are many possible ways to combine the segmentation map and the raw image, and one could even leave this task to the following regression module (e.g., using a ResNet or ViT with more than three channels as input), we deliberately choose this simple method of combination. This approach allows us to utilize regular pre-trained vision models (e.g., ResNet, ViT) when building the regression module. After combining the maps, we use the prompt to crop the regions of interest from the full image. If the prompt is provided as a bounding box, we directly use it to crop the image within that bounding box. If the prompt is given as a center point, the cropping window is determined by $\hat{y}_{i,j}$\footnote{At the current stage, we focus solely on bounding-box prompts. Point-based prompts present a more challenging task, which we plan to address in the next update.}. The overall pre-processing step is denoted as $\psi (x_i, \hat{y}_{i,j}, p_{i,j})$. An illustration of the data, both before and after pre-processing, is presented in Figure~\ref{fig:data_demo}.

\underline{Regression module:} Two main options are available for constructing the regression module: Convolution-based models and Transformer-based models. We experimented with ResNet101, ViT-b and ViT-l in the experiments. In line with the notation in~\ref{sec:overview}, the regression module is denoted as $\tau$. Overall, {$\textrm{EvanySeg}$} is denoted as $\tau (\psi (x_i, \hat{y}_{i,j}, p_{i,j}))$, where $x_i$ is an image input, $p_{i,j}$ is a given prompt, $\hat{y}_{i,j}$ is the segmentation prediction (under evaluation). The Dice score is the default evaluation metric used for generating regression targets. However, there are many different evaluation metrics, and Dice score has been shown to be limited in certain scenarios~\cite{reinke2024understanding}. To address this, we propose generating multiple metric scores. For example, we compute both Dice scores and Hausdorff distances for every training sample and then use a multi-head regression module for fitting the generated scores. This approach could involve using a ViT with two regression output heads — one for fitting Dice scores and the other for fitting Hausdorff distances.

\subsection{Model Training}
With the prepared training data, defined pre-processing ($\psi$), and regression models ($\tau$), we now proceed to describe the training process. We employ a standard multi-epoch mini-batch stochastic gradient descent (SGD) training pipeline to train the {$\textrm{EvanySeg}$}. Note that $\psi$ contains no trainable parameters, and the training process focuses solely on updating the parameters of $\tau$. For simplicity, we assume the regression model has only one output head. Let $\theta$ denote the parameters of $\tau$; model training aims to minimize the following objective with respect to $\theta$.

\begin{equation}
\min_{\theta}\sum_{m=1}^{M} \sum_{i=1}^{n}\sum_{j=1}^{k_i} loss (\tau (\psi (x_i, \hat{y}_{i,j}^m, p_{i,j})), \pi (\hat{y}_{i,j}^m, y_{i,j})).
\end{equation}

Here, $m$ denote the index for SAM variants, $i$ denote the index for images, and $j$ denote the index for objects within each image. For a given image $x_i$, $k_i$ represents the number of objects present in that image. We utilize three variants of SAM, namely, the original SAM~\cite{kirillov2023segment}, MedSAM~\cite{ma2024segment}, and SAM-Med2D~\cite{cheng2023sam}, thus setting $M$ to 3. By default, we use the mean squared error (MSE) loss as the default loss function. After training, given a test image, with prompt and the segmentation map generated by SAM or a SAM variant, the segmentation quality score is generated by $\hat{q} = \tau^{trained} (\psi (x_i^{test}, \hat{y_{i,j}^{test}}, p_{i,j}^{test}))$

\subsection{Utilities}

In this section, we describe the key applications of {$\textrm{EvanySeg}$} in medical image segmentation.

\textbf{Identifying Poorly Segmented Samples.} When deploying a segmentation model on a set of image samples, each accompanied by one or more prompts, {$\textrm{EvanySeg}$} assists in identifying low-quality segmentation maps. These flagged outputs can then be reviewed and analyzed by human experts.

\textbf{Model Comparison without Ground Truth.} When evaluating multiple segmentation models on a set of image samples with corresponding prompts, segmentations can be generated for each sample-prompt pair using all the models under consideration. {$\textrm{EvanySeg}$} can then assign a score to each segmentation, enabling a comparison of the models based on their average scores across all samples and prompts. 

\textbf{Sample-wise Model Selection at Test Time.} During testing, when multiple segmentation models are available, {$\textrm{EvanySeg}$} can be leveraged to select the most suitable model output for each individual sample. Specifically, the segmentation with the highest {$\textrm{EvanySeg}$} score is chosen as the final output for each image and prompt.


\section{Experiments and Results}
The {$\textrm{EvanySeg}$} model was trained based on 107,055 2D images, accompanied by 206,596 object-level ground truth masks\footnote{More images and masks will be used for training future versions of $\textrm{EvanySeg}$.}. Segmentation predictions for training the {$\textrm{EvanySeg}$} model were generated using SAM, MedSAM, and SAM-Med2D. In total, 619,044 object-level image-mask pairs were utilized, with each image and corresponding mask resized to $244 \times 244$ pixels for training the segmentation evaluation model. The AdamW optimizer was employed, with a mini-batch size of 128 and a learning rate of 0.0001. The primary training workload was performed using a single Nvidia Tesla A800 graphics card.

\subsection{Testing Datasets}
We utilized four public datasets (three ultrasound image datasets: TG3K~\cite{wunderling2017comparison}, TN3K~\cite{gong2023thyroid}, and DDTI~\cite{pedraza2015open}, and one CT image dataset: FLARE21~\cite{Ma2022FastAL}), along with four private in-house datasets\footnote{Since SAM and its variants were trained on large-scale public datasets, incorporating private in-house datasets ensures that both the segmentation and evaluation models are not exposed to the test samples in advance.} (one MRI image dataset, one CT image dataset, one endoscopic image dataset, and one pathology image dataset) to assess the performance of the trained {$\textrm{EvanySeg}$} model.
The TG3K dataset, derived from 16 ultrasound videos, consists of high-quality 2D ultrasound images in which the thyroid gland occupies more than 6\% of the total image area. This dataset is specifically designed for thyroid nodule segmentation and contains 3,585 images. The TN3K dataset includes 3,493 ultrasound images with detailed pixel-level annotations, split into 2,879 images for training and 614 images for testing. Meanwhile, the DDTI dataset comprises 637 ultrasound images. FLARE21 is a diverse CT dataset focused on the segmentation of four abdominal organs: the liver, spleen, pancreas, and kidneys. The dataset presents significant variability in organ shape and size, posing substantial challenges for segmentation models. FLARE21 is provided in a 3D format, from which we extracted slices to generate 2D images for model testing, yielding a total of 8,943 images.

\begin{figure}[t]
    \centering
    \includegraphics[width=0.9\linewidth]{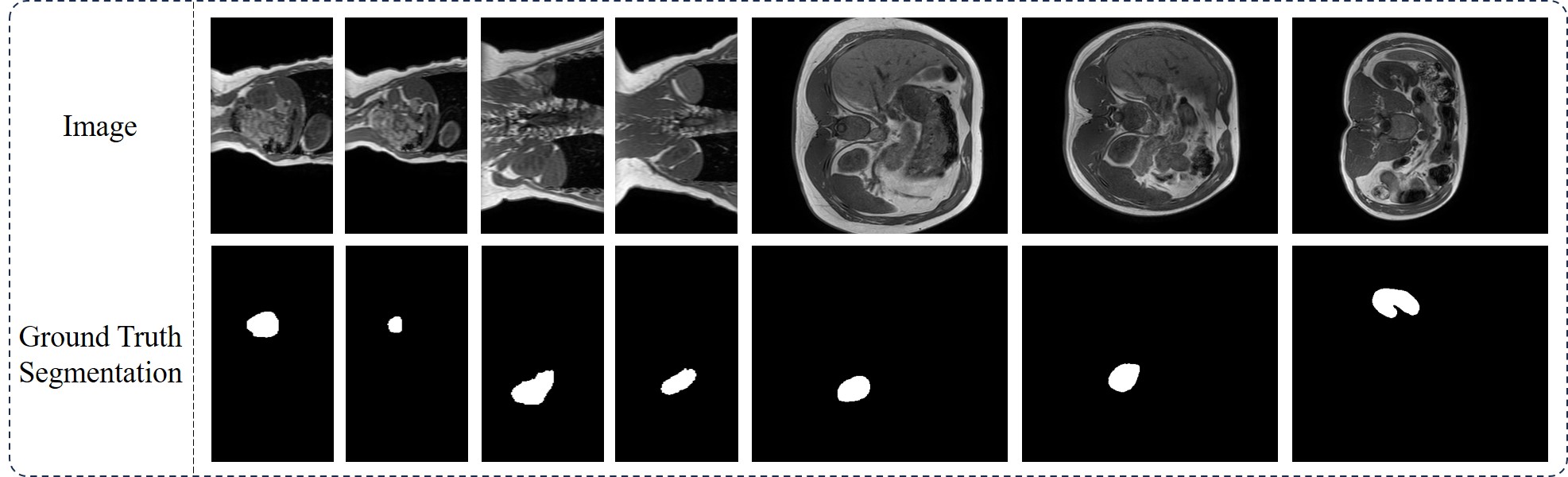}
    \caption{Samples from the MRI-Kidney100 dataset}
    \label{fig:samples MRI-kidney100}
\end{figure}

\begin{figure}[t]
    \centering
    \includegraphics[width=0.9\linewidth]{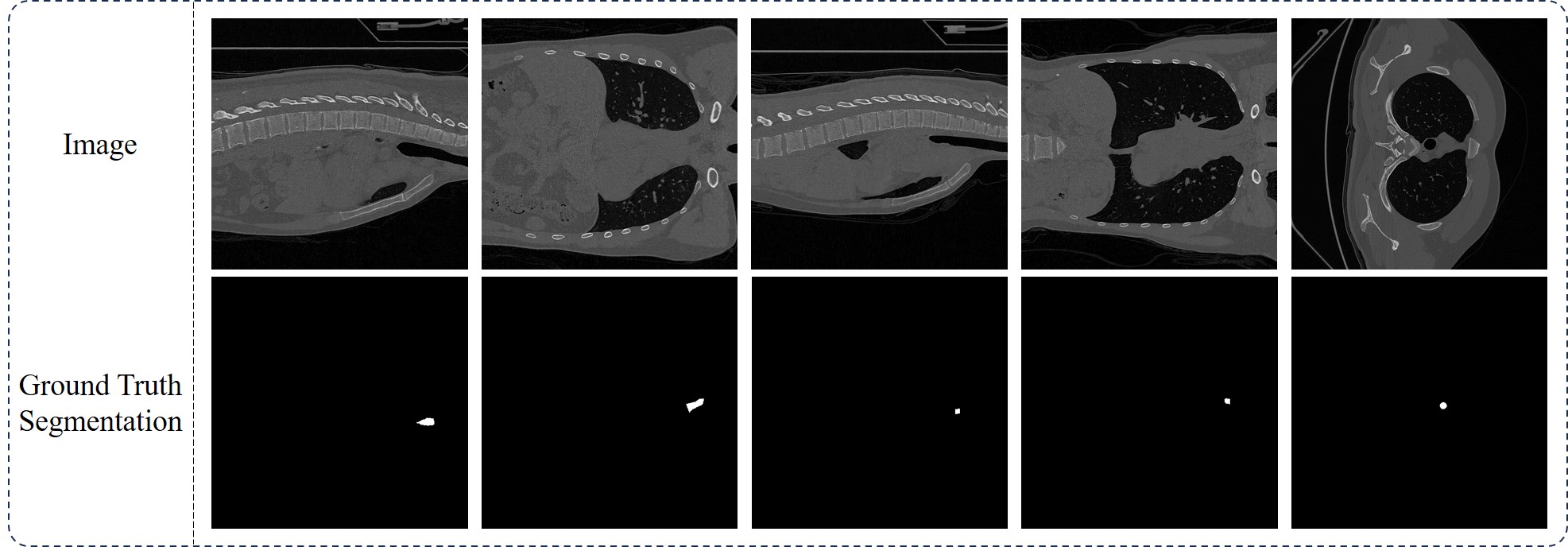}
    \caption{Samples from the CT-BCT100 dataset}
    \label{fig:samples CT-BCT100}
\end{figure}
\begin{figure}[t]
    \centering
    \includegraphics[width=0.9\linewidth]{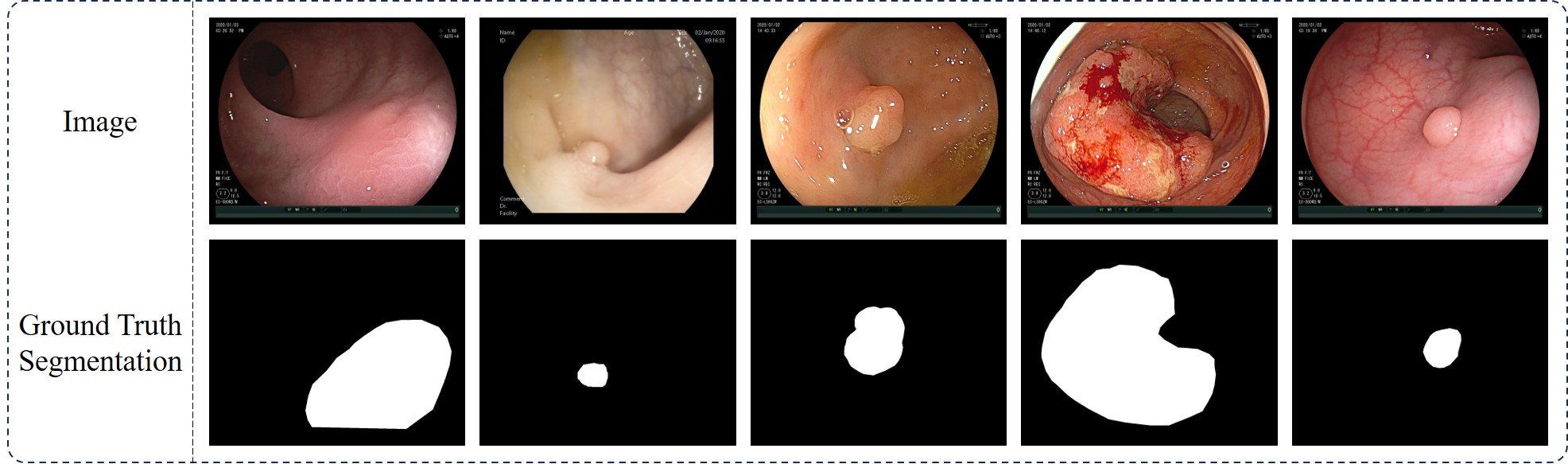}
    \caption{Samples from the Endo-Polyp1000 dataset.}
    \label{fig:PolypLarge}
\end{figure}

\begin{figure}[t]
    \centering
    \includegraphics[width=0.9\linewidth]{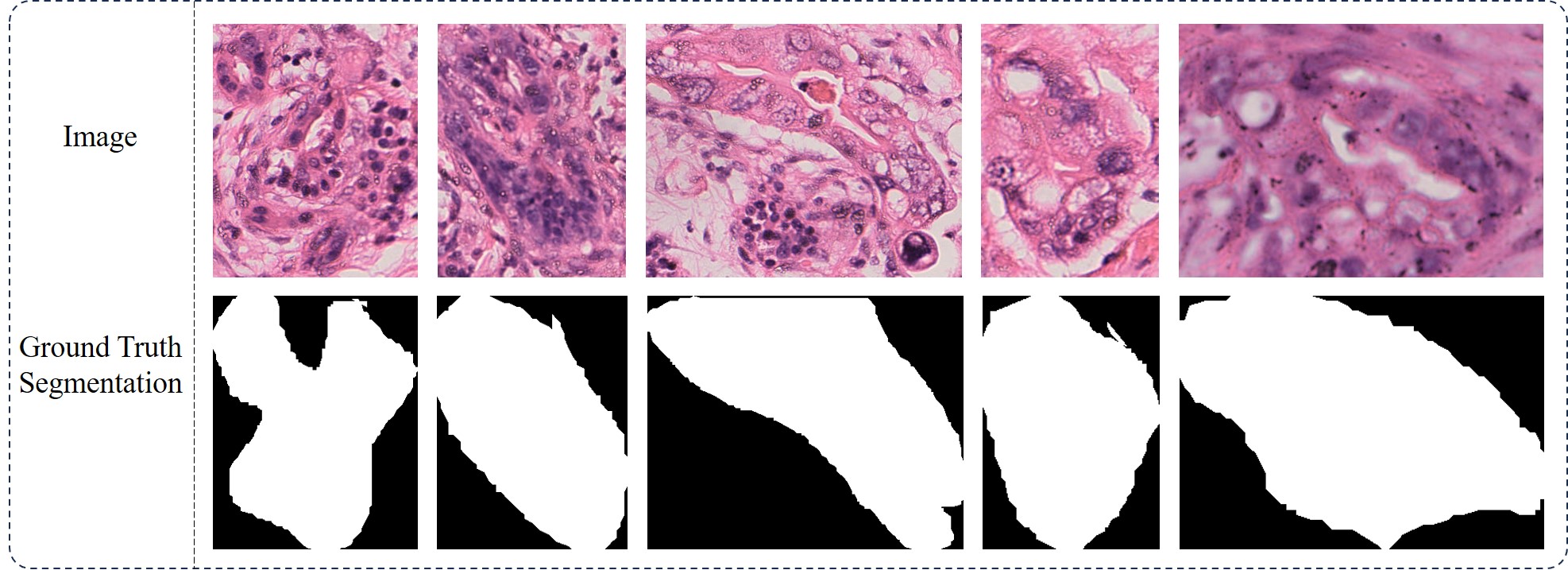}
    \caption{Cropped Samples (Regions of Interest) from the Path-Neuro1406 dataset.}
    \label{fig:PathNeuroTumor}
\end{figure}

The \textbf{1st} in-house dataset, named MRI-Kidney100, contains MRI scans from 100 individuals, designed for kidney segmentation. This dataset includes 100 3D MRI scans, from which 2D slices along the x-y plane were extracted for model testing. Similarly, the \textbf{2nd} in-house dataset, named CT-BCT100, comprises CT scans from the same 100 individuals, focused on brachiocephalic trunk (BCT) segmentation. This collection includes 100 3D CT scans, with 2D slices from the x-y plane used for model testing. The \textbf{3rd} in-house dataset, named Endo-Polyp1000, is a colonoscopy polyp detection dataset consisting of 1000 images, each paired with an expert-annotated mask. This dataset features a diverse range of polyp images. The \textbf{4th} in-house dataset, named Path-Neuro1406, is a pancreatic neuroendocrine-related segmentation dataset with 1,406 images. In this dataset, we cropped the regions of interest from the large pathological sections stained with hematoxylin and eosin (H\&E) according to doctor-annotated masks. The regions of interest vary significantly in shape and size, making segmentation particularly challenging. Image samples of these four in-house datasets can be found in Figures~\ref{fig:samples MRI-kidney100},~\ref{fig:samples CT-BCT100},~\ref{fig:PolypLarge} and~\ref{fig:PathNeuroTumor}.



\subsection{Correlation between the predicted Dice scores and the true Dice scores}\label{exp_sec1}

\begin{figure}[t] 
    \centering
    \centering
    \includegraphics[width=0.9\linewidth]{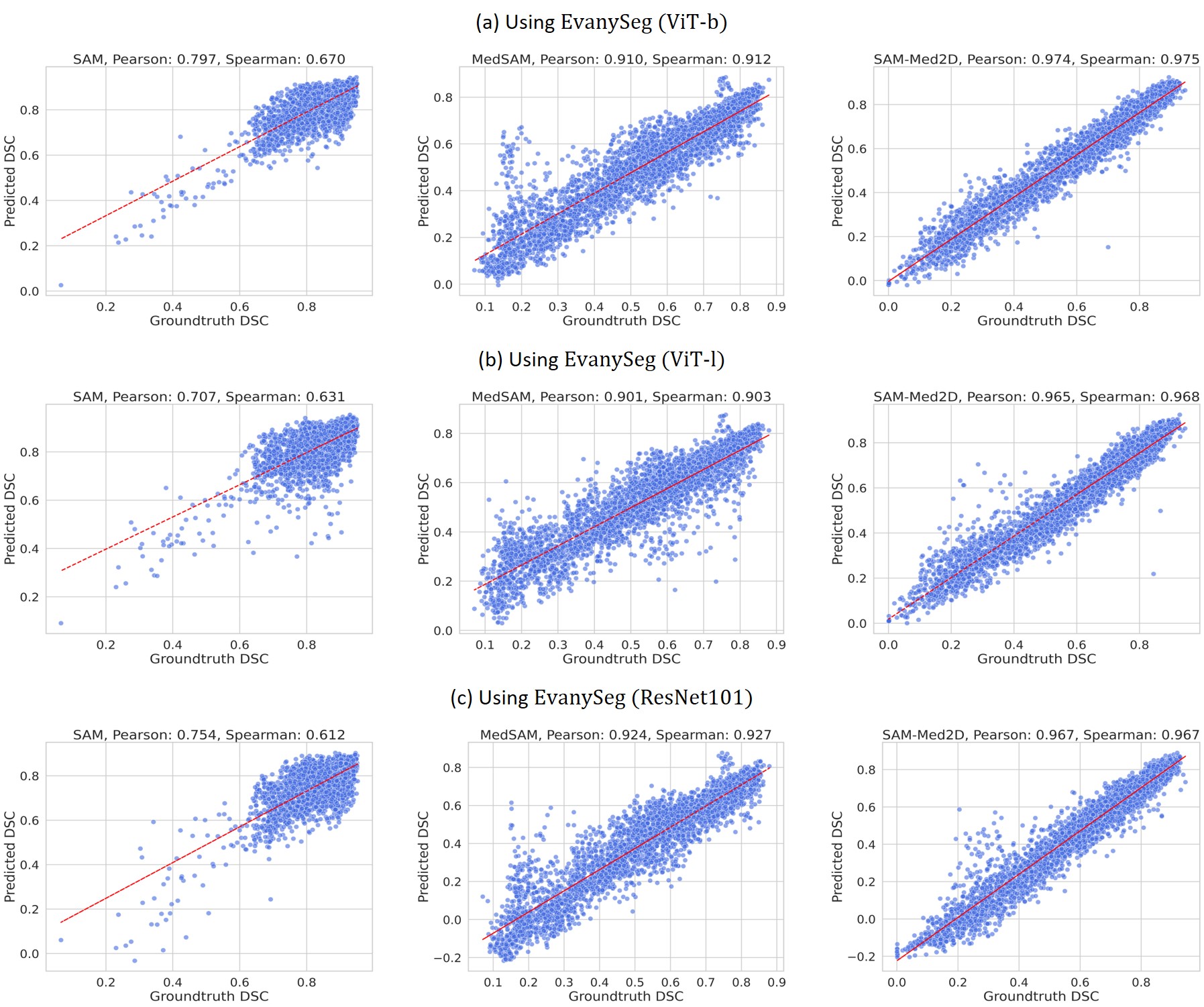}
    \caption{Tested on the TG3K dataset (Prompt Type: Bounding Box).}
    \label{fig:TG3K}
\end{figure}

\begin{figure}[h!] 
    \centering
    \includegraphics[width=0.9\linewidth]{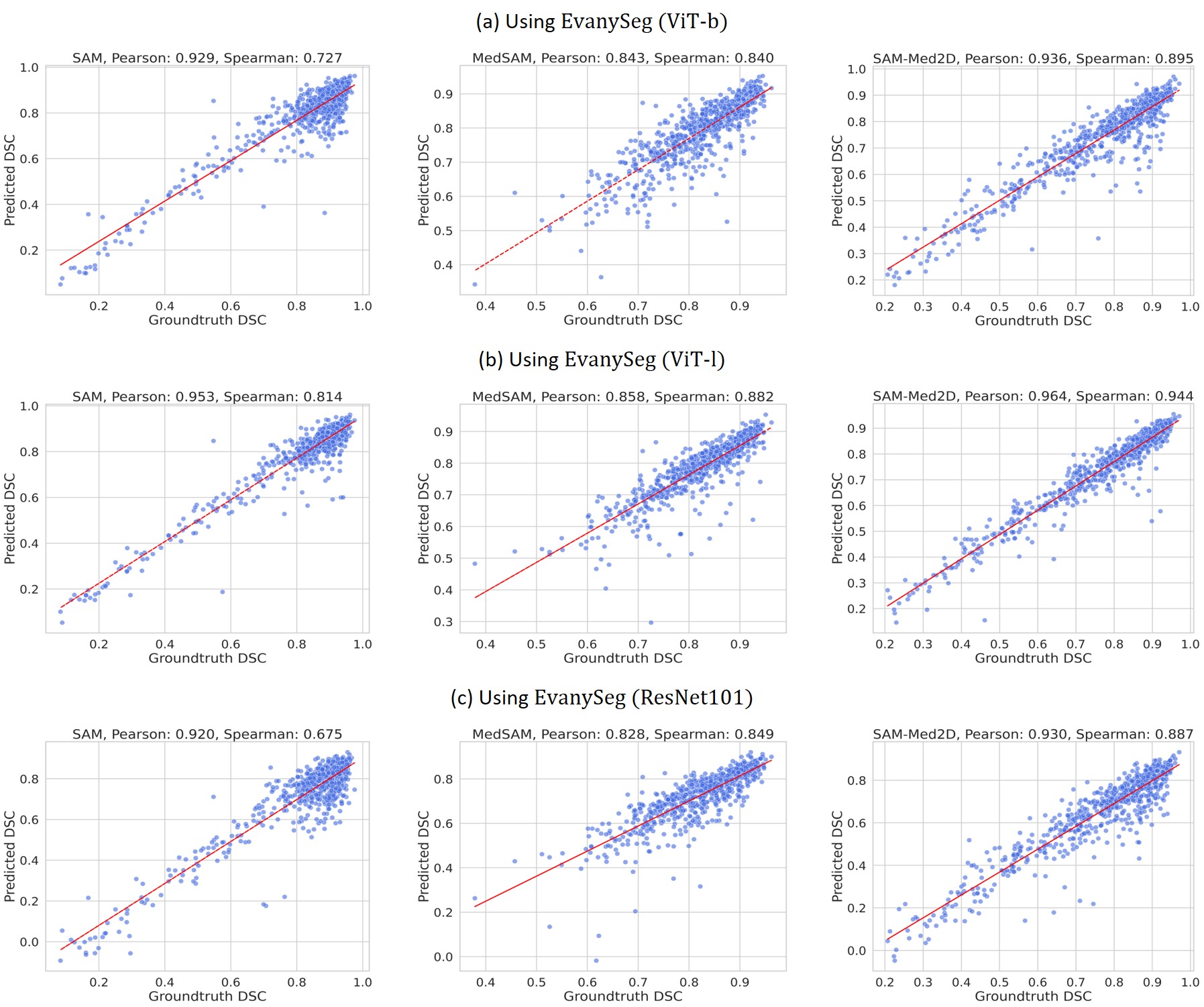}
    \caption{Tested on the TN3K dataset (Prompt Type: Bounding Box).}
    \label{fig:TN3K}
\end{figure}

\begin{figure}[h!] 
    \centering
    \includegraphics[width=0.9\linewidth]{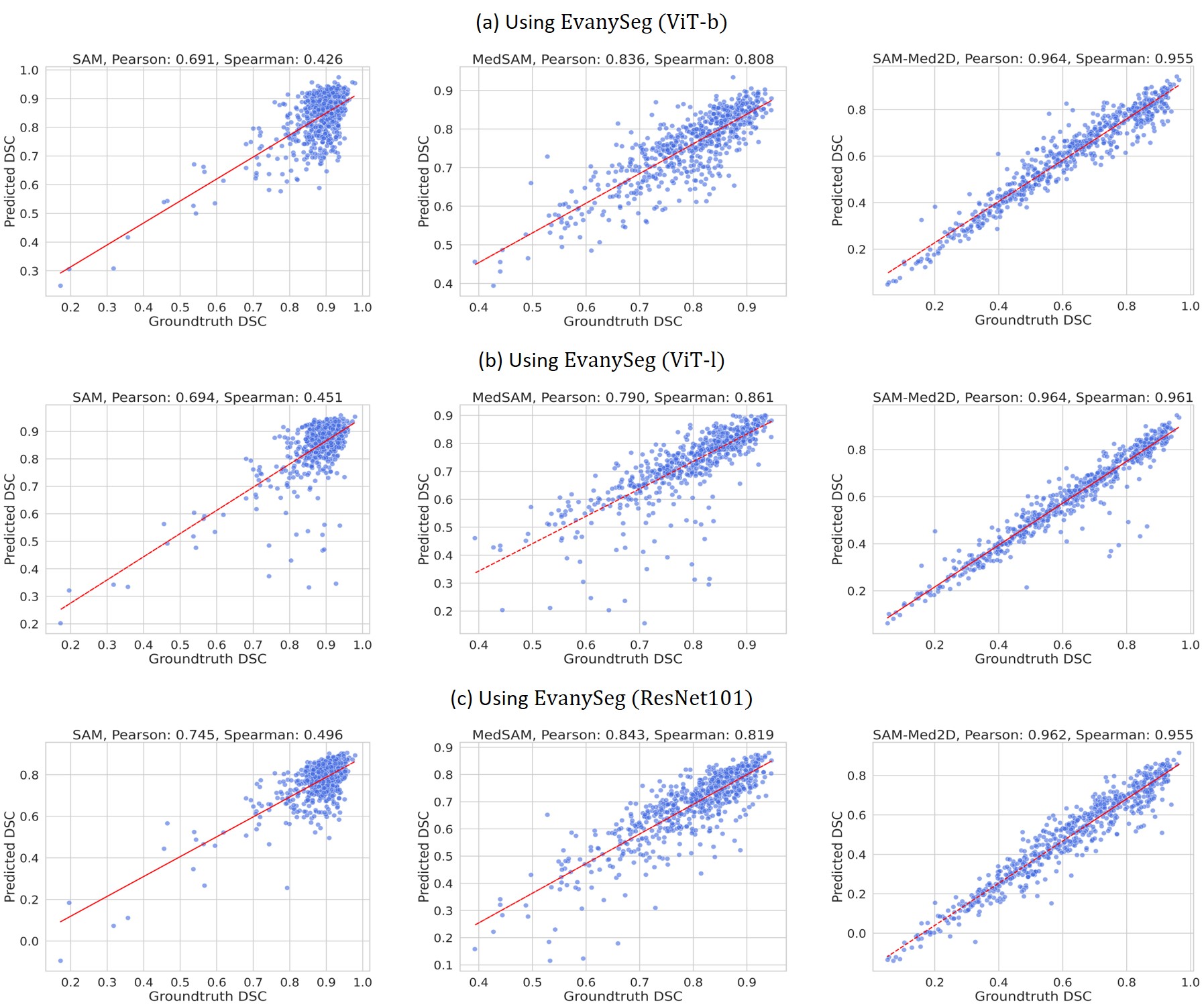}
    \caption{Tested on the DDTI dataset (Prompt Type: Bounding Box).}
    \label{fig:DDTI}
\end{figure}

\begin{figure}[h!] 
    \centering
    \includegraphics[width=0.9\linewidth]{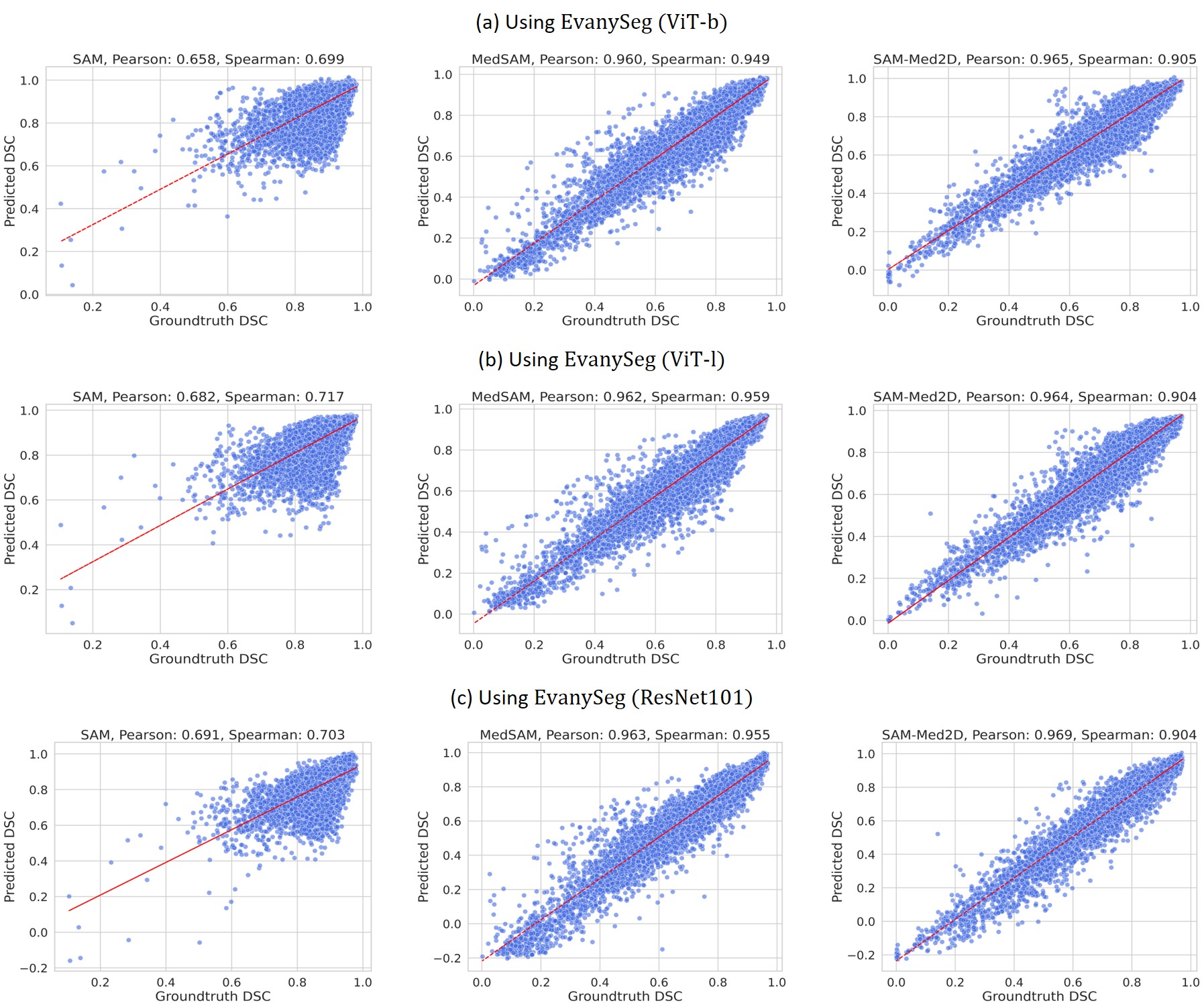}
    \caption{Tested on the FLARE21 dataset (Prompt Type: Bounding Box).}
    \label{fig:FLARE21}
\end{figure}

\begin{figure}[h!] 
    \centering
    \includegraphics[width=0.9\linewidth]{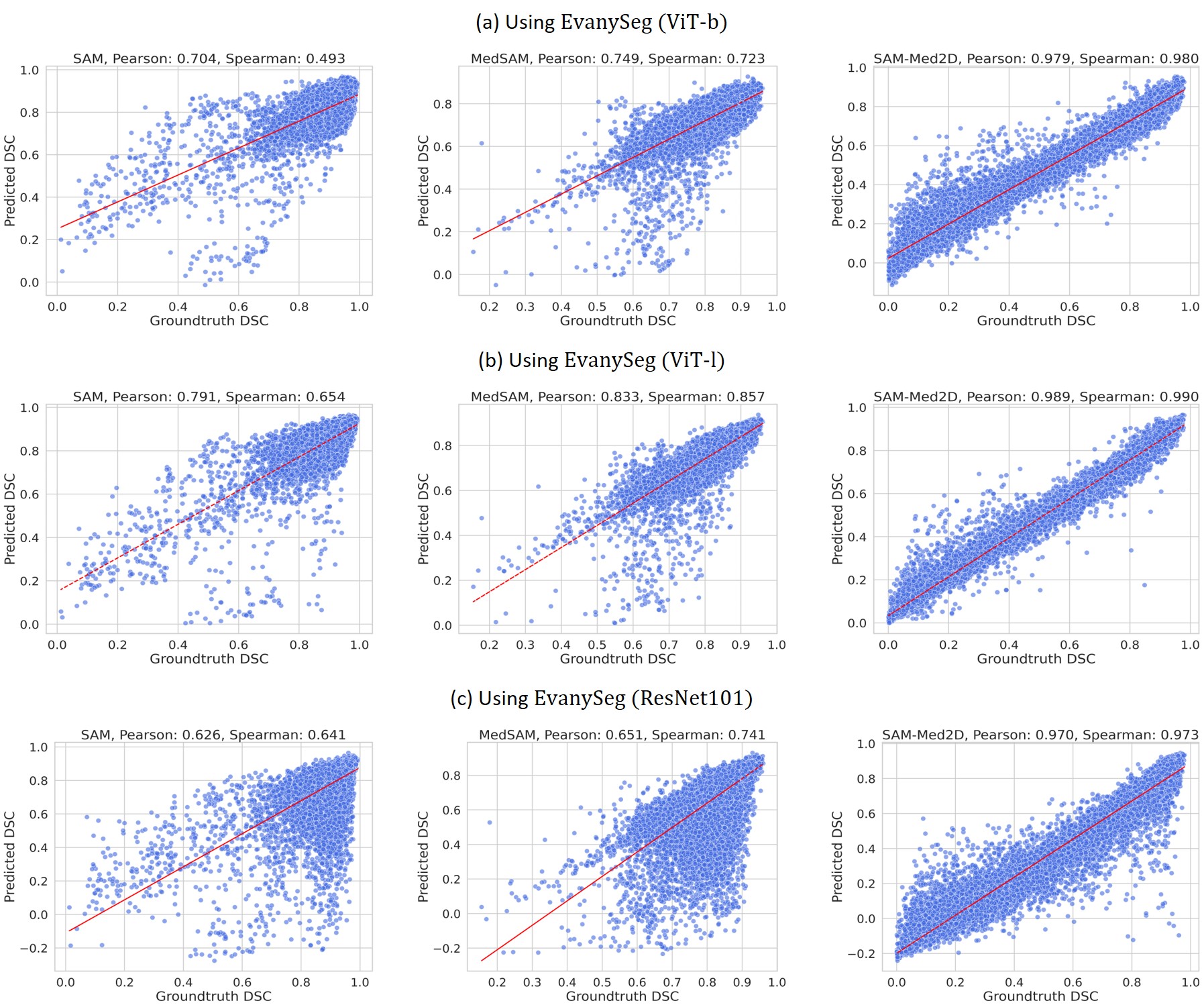}
    \caption{Tested on the MRI-Kidney100 dataset (Prompt Type: Bounding Box).}
    \label{fig:MRI-Kidney}
\end{figure}

\begin{figure}[h!] 
    \centering
    \includegraphics[width=0.9\linewidth]{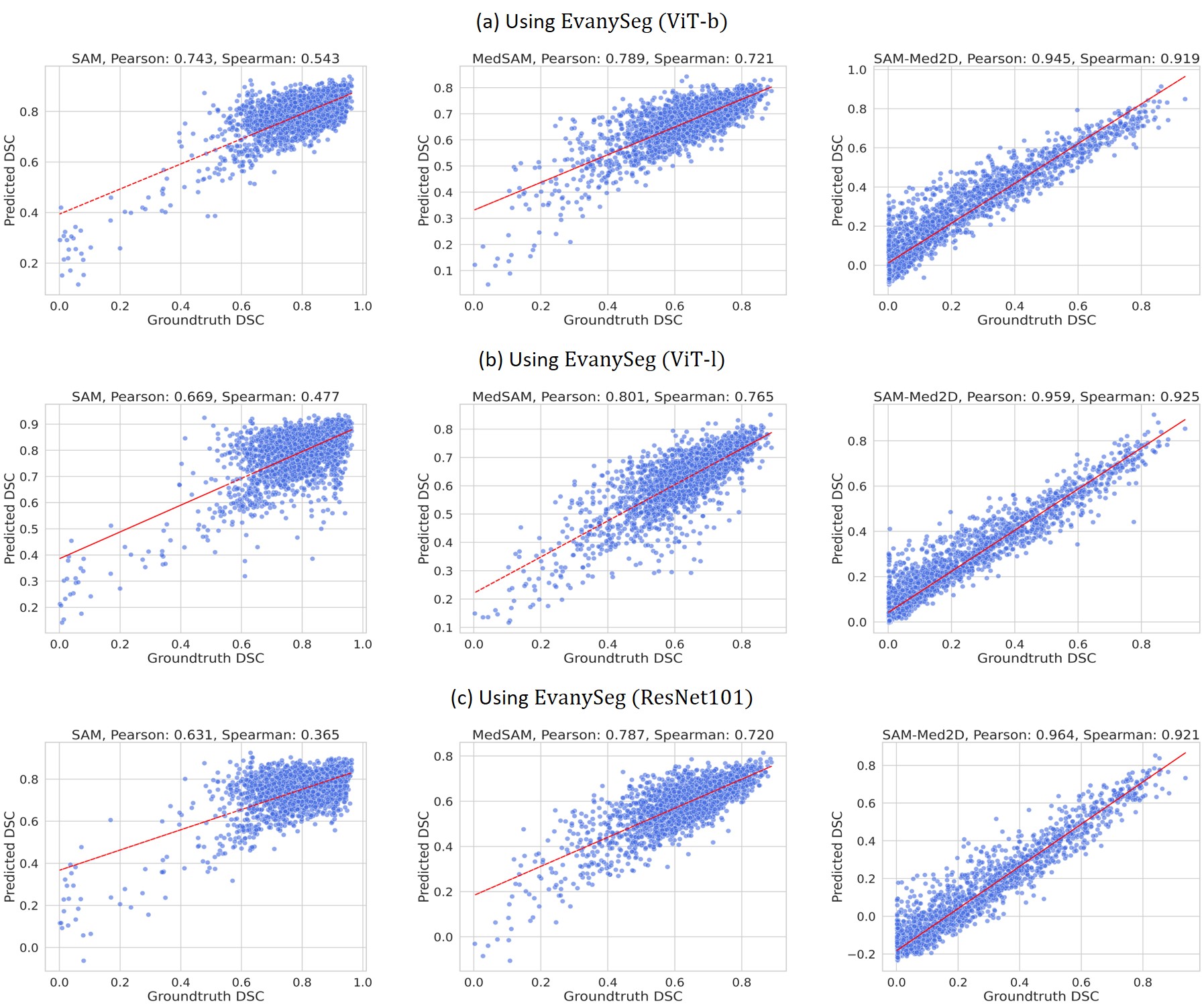}
    \caption{Tested on the CT-BCT100 dataset (Prompt Type: Bounding Box).}
    \label{fig:CT-BCT100}
\end{figure}

\begin{figure}[h!] 
    \centering
    \includegraphics[width=0.9\linewidth]{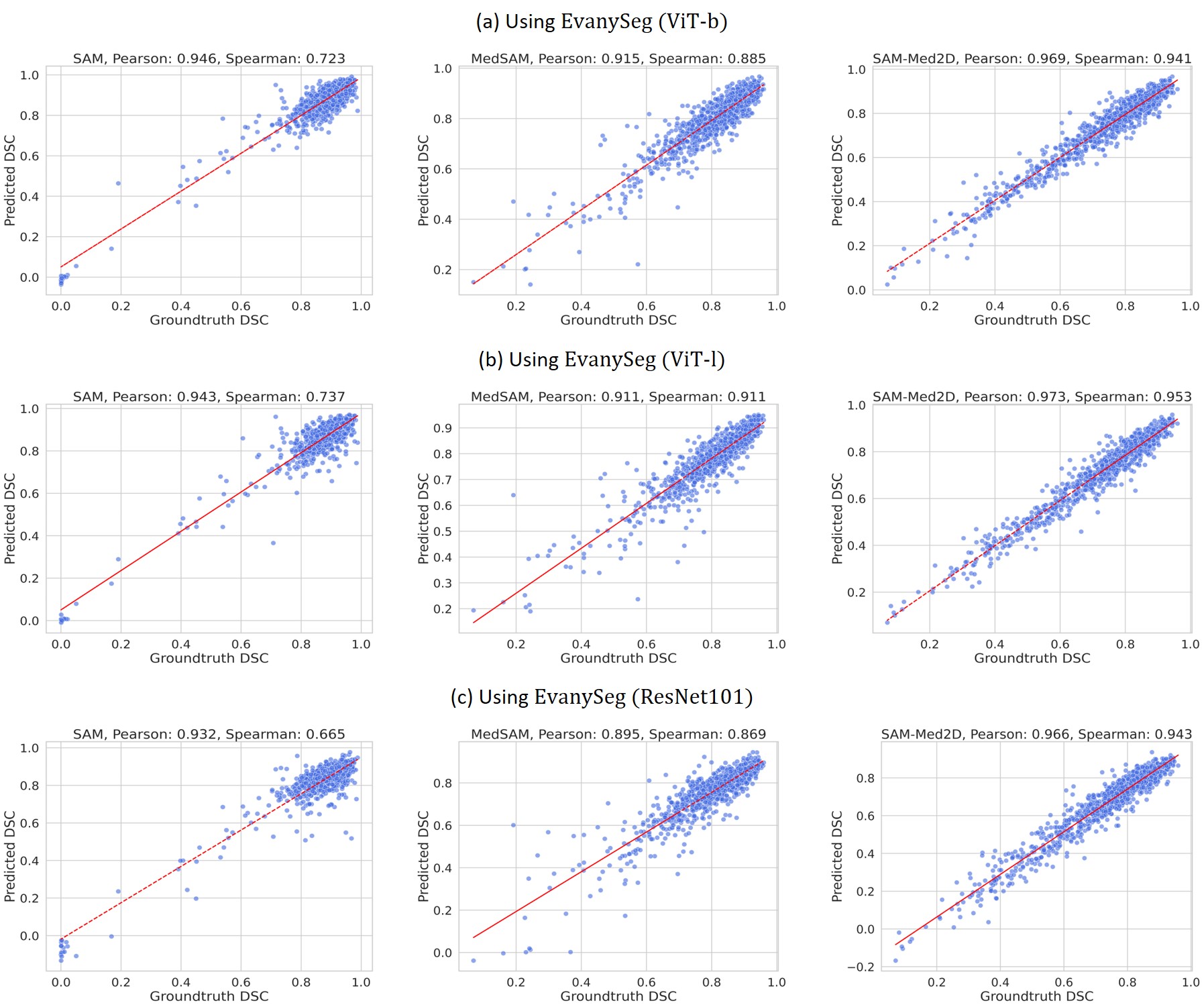}
    \caption{Tested on the Endo-Polyp1000 dataset (Prompt Type: Bounding Box).}
    \label{fig:Polyp}
\end{figure}

In this section, we evaluate how well the predicted Dice scores by {$\textrm{EvanySeg}$} correlate with the true Dice scores obtained using ground truth masks. We use Pearson correlation and Spearman's rank correlation~\cite{hauke2011comparison} for evaluation. Higher correlation scores indicate that {$\textrm{EvanySeg}$} is more accurate in predicting segmentation quality, particularly in terms of region overlaps as captured by the Dice score. Figures~\ref{fig:TG3K},~\ref{fig:TN3K},~\ref{fig:DDTI},~\ref{fig:FLARE21},~\ref{fig:MRI-Kidney},~\ref{fig:CT-BCT100}, and~\ref{fig:Polyp} present scatter plots for TG3K, TN3K, DDTI, FLARE21, MRI-Kidney100, CT-BCT100 and Endo-Polyp1000 datasets, where each point corresponds to a test image sample. Each figure corresponds to one dataset, with the first row displaying results for {$\textrm{EvanySeg}$} using ViT-b as its backbone, the second row for ViT-l, and the third row for ResNet101. The x-axis represents the true DSC (computed by comparing the segmentation prediction with the ground truth mask), and the y-axis represents the predicted DSC given by {$\textrm{EvanySeg}$} (without using ground truth masks). 

\begin{figure}
    \centering
    \includegraphics[width=1\linewidth]{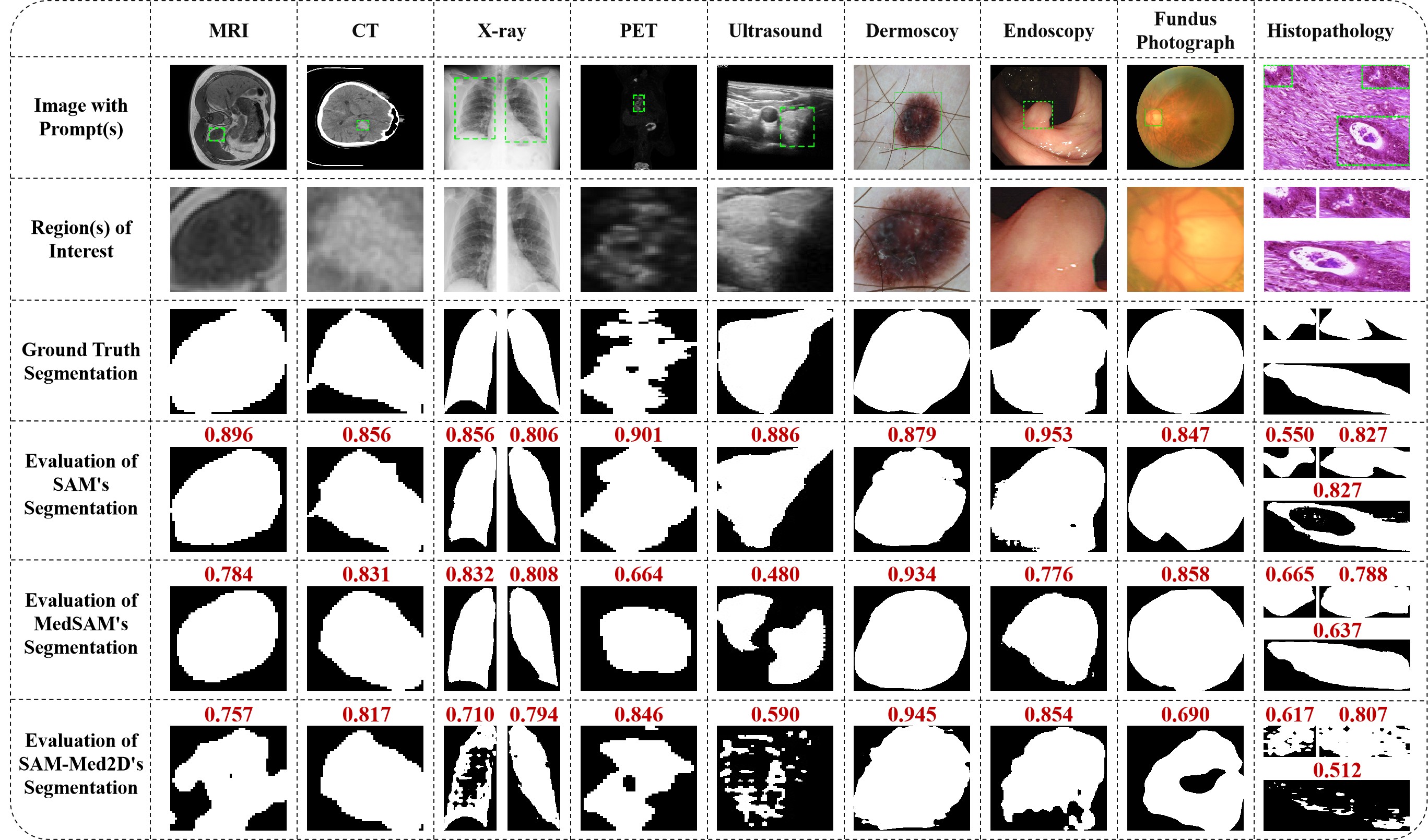}
    \caption{Visual examples of images, ground truth maps, and segmentation predictions from SAM and its variants. The red numbers are segmentation quality scores generated by {$\textrm{EvanySeg}$}.}
    \label{fig:Evaluate_segmentation}
\end{figure}

Several observations emerge from the results presented in the figures. First, ViT as a backbone, in most cases, provides better segmentation quality assessment performance compared to the ResNet counterpart. Second, segmentation predictions produced by MedSAM and SAM-Med2D exhibit significant variation in quality, while SAM generally producing better segmentation predictions than MedSAM and SAM-Med2D. Third, due to these differences in segmentation quality, predicting segmentation quality, or distinguishing between better and worse segmentation outputs, is more challenging for SAM predictions than for MedSAM and SAM-Med2D cases, as it is easier to identify poor-quality segmentations in the latter. This explains why the first columns (SAM predictions) in these figures show worse segmentation quality assessment performance than the second and third columns, as the task is generally more difficult in the SAM scenario (see Proposition~\ref{p3}). Overall, it is evident that {$\textrm{EvanySeg}$} predicts DSC with a strong correlation to the true DSC. Figure~\ref{fig:Evaluate_segmentation} presents visual examples of test images, their corresponding ground truth segmentations, and segmentation predictions from SAM and its variants. Segmentation quality scores, generated by {$\textrm{EvanySeg}$}, are provided for visual inspection of how well the scores reflect actual segmentation quality.

\subsection{Evaluation of Segmentation from Non-SAM Models}
We employ a task-specific segmentation model that is not from the SAM family and is not part of the training process of {$\textrm{EvanySeg}$} to generate segmentation predictions for testing {$\textrm{EvanySeg}$}. In Table~\ref{tab:PathNeuroTumor}, we present the correlation results using the Path-Neuro1406 dataset for this configuration. This dataset presents notable challenges, including the irregular shapes of objects, significant size variations, and the lack of clear visual boundaries in the H\&E-stained histology images. The ``ALL Samples" column in Table~\ref{tab:PathNeuroTumor} shows the correlations between the predicted Dice scores from {$\textrm{EvanySeg}$} and the actual Dice scores. The ``Large Object" and ``Small Object" columns detail the correlation results for {$\textrm{EvanySeg}$} with respect to larger and smaller objects.

\begin{figure}[h!] 
    \centering
    \includegraphics[width=0.9\linewidth]{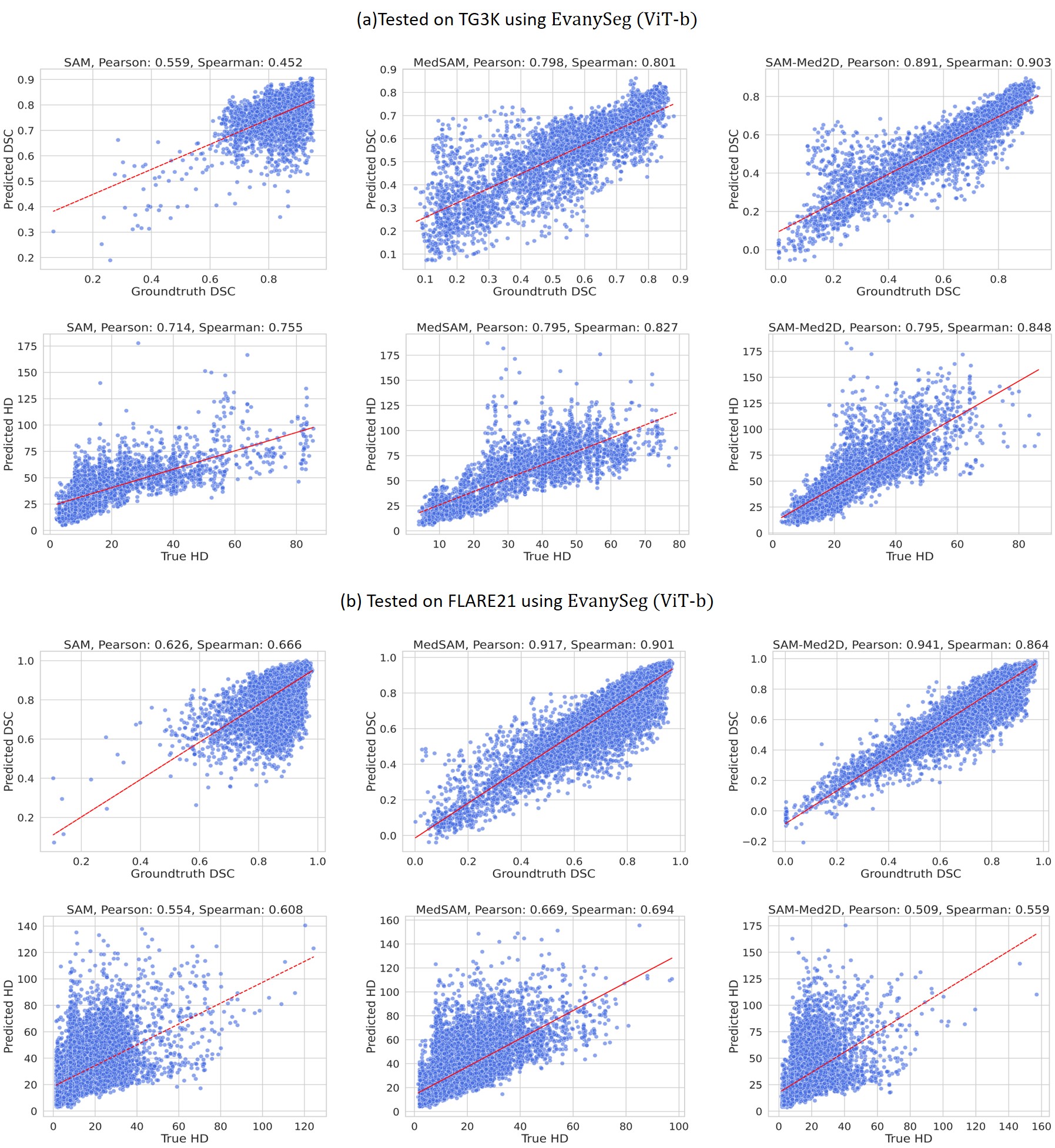}
    \caption{{$\textrm{EvanySeg}$}  with two prediction heads: one for predicting Dice score and the other for predicting HD (Hausdorff Distance), using a bounding box prompt.}
    \label{fig:dice_HD}
\end{figure}




\begin{table}[h!] 
     \centering
     \caption{{Performance of $\textrm{EvanySeg}$} in predicting DSC on the in-house Path-Neuro1406 dataset for the segmentation predictions generated by a task-specific segmentation model (not a SAM).}
     \begin{tabular}{|c|c|c|c|c|}
        \hline
         Backbone&Correlation & All Samples & Large Object  & Small Object \\ 
         \hline
         \multirow{2}{*}{ViT-b} &Pearson & 0.506& 0.563 &0.406 \\
         \cline{2-5}
         &Spearman &0.516 &0.604  &0.416 \\
         \hline
         \multirow{2}{*}{ResNet101}&Pearson &0.469  & 0.467 & 0.429\\
         \cline{2-5}
         &Spearman & 0.504 &0.502  &0.487 \\
         \hline
     \end{tabular}
     
     \label{tab:PathNeuroTumor}
 \end{table}

\begin{table}[h!] 
    \centering
    \caption{Accuracy(\%) in sample-wise selecting the best segmentation predictions (among SAM, MedSAM and SAM-Med2D.)}
    \begin{tabular}{|c|c|c|c|}
        \hline
        Dataset  &   {$\textrm{EvanySeg}$(ViT-b)}    & {$\textrm{EvanySeg}$(ViT-l)}  & 
        {$\textrm{EvanySeg}$(ResNet-101)}\\
        \hline
         TG3K    &  90.83  &  89.38  &  89.86  \\
         \hline
        FLARE21  &  77.00  &  77.49  &  75.71  \\
        \hline
    \end{tabular}

    \label{tab:SMS-1}
\end{table}

\begin{table}[h!] 
    \centering
    \caption{Left panel: Performance (Dice score) of sample-wise model selection. Right panel: Performance of each segmentation model (candidate model for model selection). The results of the ``w/ Oracle'' indicate the performance upper bound (for the selection part).}
    \begin{tabular}{|c|c|c||c|c|c|}
    \hline
        Dataset  &  w/ Oracle &w/ {$\textrm{EvanySeg}$(ViT-b)} & SAM & MedSAM &SAM-Med2D \\
            \hline
         TG3K   &85.6 & 85.2  & 84.6 & 51.5& 51.6   \\
             \hline
        FLARE21  &91.4& 90.6  & 89.2 & 74.4 & 74.3    \\
            \hline
    \end{tabular}

    \label{tab:SMS-2}
\end{table}

\subsection{Multiple Regression Outputs for Multiple Evaluation Metrics}
We further experimented with {$\textrm{EvanySeg}$}, where the model is designed with two output heads—one for predicting Dice scores and the other for predicting Hausdorff distances. In Figure~\ref{fig:dice_HD}, we present the scatter plots and computed correlation values for the two-head configuration. Compared to the single-head setup in Figure~\ref{fig:TG3K} and~\ref{fig:FLARE21}, we observe a slight decrease in correlation values (for Dice score prediction) when using the two-head regression, where the Hausdorff distance predictions share weights in the ViT architecture with the Dice score predictions. This observation highlights the need for further architectural improvement of {$\textrm{EvanySeg}$} to enable more effective and efficient multi-task learning.

\subsection{Using {$\textrm{EvanySeg}$} for Sample-wise Model Selection}\label{exp_sec2}
We then evaluate whether {$\textrm{EvanySeg}$} can accurately determine which segmentation model performs best for a given test image when multiple segmentation models are available. The experiments are conducted using the TG3K and FLARE21 datasets, employing the predicted scores and true Dice scores from Section~\ref{exp_sec1}. Table~\ref{tab:SMS-1} demonstrates that {$\textrm{EvanySeg}$} correctly identifies the best model (among SAM, MedSAM, and SAM-Med2D) approximately 90\% of the time on the TG3K dataset, and about 77\% of the time on the FLARE21 dataset. In Table~\ref{tab:SMS-2}, we further demonstrate that using {$\textrm{EvanySeg}$} as a sample-wise model selector improves segmentation performance (measured by Dice score) compared to scenarios where only a single model is applied without selection. We also present results for the case where the Oracle selector is deployed, representing the upper-bound performance limit (for the selection part).

\section{Discussion}
Data-driven and deep learning approaches have revolutionized the field of medical image segmentation over the past decade. From U-Net~\cite{ronneberger2015u} to nnU-Net~\cite{isensee2021nnu}, and now to Transformer-based models~\cite{xiao2023transformers}, segmentation techniques have continually advanced through the integration of larger and better datasets, improved model architectures, additional prior knowledge, and even multi-modal modeling, including text-based information. Despite these advancements, challenges remain in ensuring the reliability and trustworthiness of the most advanced segmentation models when deployed in clinical practice. Accurately and robustly evaluating segmentation quality without relying on ground truth data is crucial for the reliable deployment of medical AI models. When an evaluator detects low-quality segmentation output, the system can alert human experts to examine the corresponding sample and subsequent results. In this paper, we successfully developed such a segmentation evaluation model as a companion to the Segment Anything Model for medical images. Moreover, we demonstrate that this evaluator can be employed to select the best model for each test sample during test time, thereby actively improving segmentation outcomes.

Our approach to building a segmentation evaluation model is not particularly novel: The use of deep learning and regression objectives to estimate Dice scores has been previously studied (e.g.,~\cite{robinson2018real}). Similarly, the application of Vision Transformers (ViT) for visual learning in medical data has been extensively explored (e.g.,~\cite{chen2021transunet}), and the objective of creating a universal segmentation evaluator has also been proposed in a very recent study (\cite{chen2024quality}). Our focus is in developing a companion segmentation evaluation model specifically for the Segment Anything Model applied to medical data, which is a new advancement in this domain. We deliberately chose well-established options in model architecture, training methods, and data preparation to demonstrate that, even with a straightforward setup, it is possible to achieve a reasonably effective evaluator for ground-truth-free segmentation quality assessment. We believe that with improvements such as a larger model with a more advanced architecture, better dataset preparation (e.g., more high-quality data), and an enhanced training pipeline (e.g., with data augmentation), {$\textrm{EvanySeg}$} could achieve even better performance in estimating scores that more closely align with true segmentation quality measures.

\textbf{Limitations.} First, despite the high correlation observed between the predicted DSC and true DSC in the experiments and results, a closer examination at a very local level would reveal that the order of the predicted Dice scores can still be noticeably incorrect relative to the true Dice scores (the problem is harder at this level, see Proposition~\ref{p3}). Secondly, {$\textrm{EvanySeg}$}, as a companion model to SAM, shares SAM's limitation of relying on prompts (e.g., points and/or bounding boxes) for generating segmentation. In some cases, the segmentation which should be generated from a bounding-box or point prompt—especially with point prompts—is not well-defined. Thirdly, {$\textrm{EvanySeg}$} currently works only with 2D medical images, and for the present version of the manuscript, it only supports bounding-box prompts. Point-based prompts will be addressed in the next update.

\section{Conclusion}
In this paper, we introduce a segmentation evaluation model designed to complement SAM-type segmentation models, serving as a companion tool for assessing segmentation quality without the need for ground truth masks. Theoretical analysis confirms the viability of this approach, while the development process emphasizes simplicity and clarity, highlighting its potential. Experimental results demonstrate the model’s effectiveness, with the generated scores aligning closely with true segmentation quality. We showcase several practical uses for these scores, such as identifying poor segmentations, selecting superior models, and optimizing segmentation outcomes at the sample level across multiple models. Through this paper, along with our code and trained models, we aim to promote the creation of more accurate, faster, and robust ground-truth-free or ground-truth-less-dependent segmentation evaluation tools, advancing research in trustworthy medical AI and enhancing human-AI collaboration in medical image analysis workflows.

\bibliographystyle{alpha}
\bibliography{sample}

\end{document}